\documentclass[prb,aps,twocolumn,groupedaddress,floats,showpacs,final]{revtex4-1}
\usepackage{graphicx}
\usepackage{dcolumn}
\usepackage{bm}
\usepackage{color}
\definecolor{blue}{rgb}{0.3,0.3,0.9}

\begin{document}

\author{Lode Pollet}
\affiliation{Department of Physics, Arnold Sommerfeld Center for Theoretical Physics and Center for NanoScience, University of Munich, Theresienstrasse 37, 80333 Munich, Germany}
\author{Nikolay V. Prokof'ev}
\affiliation{Department of Physics, University of Massachusetts,
Amherst, MA 01003, USA}
\affiliation{Russian Research Center ``Kurchatov Institute'',
123182 Moscow, Russia}
\author{Boris V. Svistunov}
\affiliation{Department of Physics, University of Massachusetts,
Amherst, MA 01003, USA}
\affiliation{Russian Research Center ``Kurchatov Institute'',
123182 Moscow, Russia}


\title{Asymptotically Exact Scenario of Strong-Disorder Criticality in One-Dimensional Superfluids}


\date{\today}
\begin{abstract}
We present a controlled rare-weak-link theory of the superfluid-to-Bose/Mott glass transition
in one-dimensional disordered systems. The transition has Kosterlitz-Thouless critical properties but may
occur at an arbitrary large value of the Luttinger parameter $K$.
 The hydrodynamic description is valid under the correlation radius and defines criticality  via mutual renormalization of
the strength of microscopic weak links and  superfluid stiffness. The link strength renormalizes along the lines of Kane and Fisher [Phys. Rev. Lett. {\bf 68}, 1220 (1992)],
while the renormalization of superfluid stiffness follows the lines of classical-field flow.
The hallmark of the theory is the relation $K^{(c)}=1/\zeta$ between the critical value
of the Luttinger parameter at macroscopic scales and the microscopic
(irrenormalizable) exponent $\zeta$ describing the scaling $\propto 1/N^{1-\zeta}$
for the strength of the weakest link among the $N \gg L$ disorder realizations
in a system of fixed mesoscopic size $L$.

\end{abstract}
\pacs{03.75.Hh, 67.85.-d, 64.70.Tg, 05.30.Jp}

\maketitle

\section{Introduction}
\label{sec:intro}

Scalar bosons with local interactions in one dimension are generically described by the  paradigm of Luttinger liquids (LL),  which amounts to quantized superfluid hydrodynamics augmented with instantons (aka ``backscattering events" in the fermionic language) responsible for quantum slippages of the superfluid phase. It is via the instantons that superfluid hydrodynamics is coupled to either a commensurate external potential, or disorder, or both  (see, e.g., Refs.~\onlinecite{Kashurnikov96,Svistunov96}). The LL picture is typically preserved under the correlation length
of the superfluid-to-insulator quantum phase transition, providing a natural framework for an asymptotically
exact description of criticality.

Arguably, the most intriguing superfluid-to-insulator quantum phase transition is the one that is induced by disorder and leads to the formation of the Bose glass (BG), a compressible insulator.\cite{GS,Fisher}
In their seminal paper on localization in one-dimensional (1D) superfluids,\cite{GS}  Giamarchi and Schulz found---by means of a perturbative renormalization group (RG) analysis---that the transition to the BG is of
the Kosterlitz-Thouless (KT) type and is characterized by the universal value $K^{(c)} = 3/2$ of the Luttinger parameter $K$. Recently, this finding was shown to hold at the two-loop level,\cite{Ristivojevic} in line with the  earlier proof\cite{Kashurnikov96} that $K^{(c)} = 3/2$ is a generic property rather than a weak-disorder limiting case.

The situation with strong disorder, on the other hand, remains controversial. Altman {\it et al.} conjectured\cite{Altman} that power-law distributed weak links can lead to
a non-universal value of $K^{(c)}$ (see also recent Ref.~\onlinecite{Pielawa}). To corroborate their idea, the authors attempted a scenario in which they abandoned the usual hydrodynamic description in favor of the ``Coulomb blockade" nomenclature allowing them to apply a real-space RG treatment. However, the approach is not asymptotically exact and, as we show below, inconsistent with  hydrodynamics in the superfluid state (SF). Recently, we argued\cite{classical} that the only possible effect of strong disorder is a prolonged classical flow based on the vanishing fugacity of weak links. We also proved a theorem of critical self-averaging implying that the LL picture should hold at criticality. Based on the classical-flow picture and the above-mentioned theorem, we claimed no alternative to the  Giamarchi-Schultz universality class.

In the present work, we observe that our Ref.~\onlinecite{classical} contains an arbitrary statement saying that
applicability of hydrodynamics  necessarily implies Giamarchi-Schultz criticality, along with a major flaw: the quantum hydrodynamic renormalization of weak links was overlooked.
However, if the flaw is corrected,  an asymptotically exact theory of a new universality class of superfluid to
Bose/Mott glass transition in one dimension emerges, which we derive below.

 The paper is organized as follows. In Sec.~\ref{sec:KF}, we introduce the basic notation for the hydrodynamical description and review the RG flow of a single weak link in a homogeneous superfluid. In Sec.~\ref{sec:classical}, we render the RG description of the classical-field flow in the presence of strong disorder, following Ref.~\onlinecite{classical}. We then argue that these two flows must be combined, which intuitively yields the central result of the paper, Eq.~(\ref{crit_cond}). Consequently, we rigorously prove Eq.~(\ref{crit_cond}) in Sec.~\ref{sec:semiRG}, where asymptotically exact  semi-RG flow equations are derived. A technically involved---but quite important for justifying the theory---aspect, that the relevant weak links are microscopic and isolated from each other, is referred to Sec.~\ref{sec:coulomb}. In the Conclusion (Sec.~\ref{sec:conclusion}), we summarize the main results and argue that, despite the existence of two universality classes for the transition from
a superfluid to Bose (Mott) glass, only one Bose (Mott) glass phase exists.

\section{A single weak link in a Luttinger liquid}
\label{sec:KF}

The starting point for the theoretical analysis of liquids in (1+1)d is Popov's hydrodynamic action over the phase field $\Phi(x, \tau)$,
\begin{equation}
S[ \Phi] =\int dx d\tau \left[ i n_0(x) \Phi_{\tau}'+ \frac{\Lambda_s}{2} (\Phi_x')^2 + \frac{\kappa}{2} (\Phi_{\tau}')^2 \right] \:.
\label{Popovaction}
\end{equation}
Here, $x$ stands for the spatial coordinate and $\tau$ for the imaginary time.
The fields $\Phi_{\tau}'$ and $\Phi_{x}'$ are the derivatives of the field $\Phi$ with respect to $\tau$ and $x$, respectively.
The quantity $n_0(x)$ denotes the local number density where the spatial dependence originates from an external potential. The shape of the first term shows that it is of topological nature and is able to destroy superfluidity. The Luttinger parameter $K = \pi \sqrt{\Lambda_s \kappa}$ is directly related to the compressibility $\kappa$ and the superfluid stiffness $\Lambda_s$.

\begin{figure}[htbp]
\includegraphics[scale=0.4,angle=0,width=0.8\columnwidth]{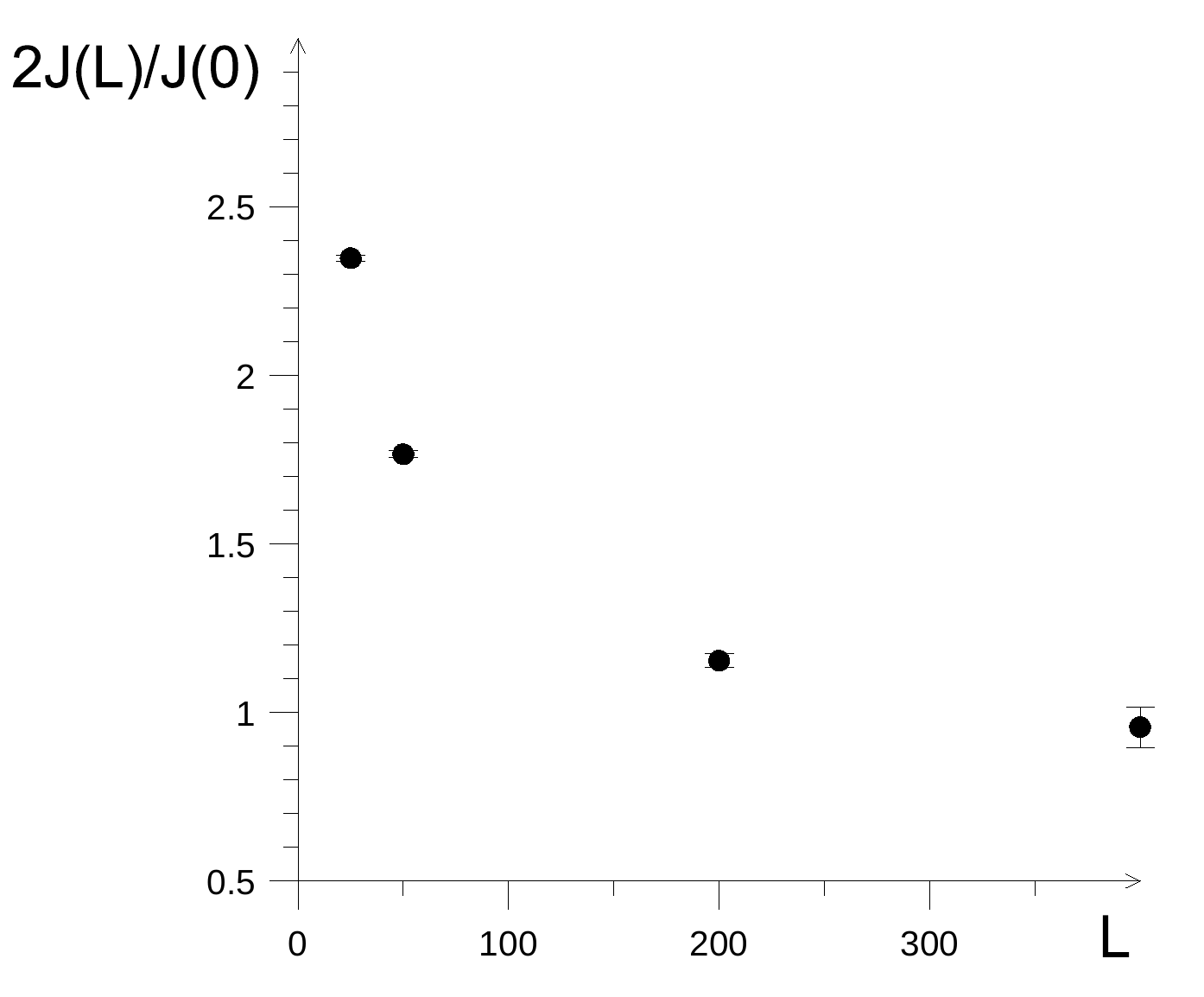}
\caption{\label{fig:onelink} Renormalization of the effective link strength
for $K\approx 3.9$ when one hopping amplitude is suppressed by a macroscopic factor
$2/L$ to form a weak link. The plot is for $2J(L)/J(0)=[\Lambda_s^{-1}(L) -\Lambda_s^{-1}(0)]^{-1}$.
While our theory for disordered systems can perfectly describe this Kane-Fisher renormalization, the real-space RG approach~\cite{Altman} can not.
}
\end{figure}

In their seminal work,\cite{KaneFisher} Kane and Fisher addressed the question of  renormalization of the strength of a weak link
in an otherwise homogeneous Luttinger liquid, finding, in particular, that, irrespectively of the microscopic strength of the link,
the superfluid transport gets completely blocked at $K<1$. The hopping across a link of strength $J_0$ is described by the term
\begin{equation}
J_0\int d\tau \cos \left[ \Phi_+(\tau) - \Phi_-(\tau) \right]
\label{wl}
\end{equation}
added to the hydrodynamic action.
Here, $\Phi_\pm $ are the values of the (1+1)-dimensional phase field $\Phi (x,\tau )$ right before and
after the weak link.  The link can be considered weak if its strength $J_0$ satisfies
the condition
\begin{equation}
J_0 \, \ell_0 \, \ll\,  \Lambda_s^{(0)} \, ,
\label{wl_cond}
\end{equation}
where $\ell_0$ is the length scale of the ultraviolet cutoff for the hydrodynamic description and  $\Lambda_s^{(0)}$  the superfluid stiffness at the scale $\ell_0$. In what follows, we measure all lengths in units of $\ell_0$ and the stiffness in units of $\Lambda_s^{(0)}$.  We note in passing that a large value of $K_0=K(\ell_0)\gg 1$ then implies a large compressibility $\kappa_0$.

The full problem posed by Eq.~(\ref{wl_cond}) is non-trivial (but solved long ago and well
understood). Progress can be made by using the renormalization group and noting that such a weak link is perturbative with respect to the short-wave harmonics of $\Phi$.
Therefore, those harmonics can be eliminated from the theory\cite{KaneFisher} by averaging  $\langle \, \ldots \, \rangle_{\lambda}$ over  harmonics with wavelength shorter than $\lambda$ and resulting in the corresponding
renormalization of the link strength, $J_0 \to J(\lambda)$, where $\lambda$ is the new ultraviolet cutoff (in units of $l_0$, in accordance with the above-mentioned units).
From the scaling dimension of the operators it follows that this flow can be written as
\begin{equation}
\frac{d J(\lambda)}{d \ln \lambda} = -\frac{1}{K} J(\lambda).
\label{eq:RGflow_KF}
\end{equation}
In our subsequent analysis for disordered systems, it will be more economical to use an integral formulation of flow equations. Let us therefore re-express the Kane-Fisher renormalization of weak links in this language. Averaging over the harmonics of $\Phi$ with wavelength shorter than $\lambda$ leads to
\begin{equation}
\langle \, \cos (\Phi_+ - \Phi_-) \, \rangle_{\lambda} \, = \, \lambda^{-1/K} \cos \left( \Phi_+^{(\lambda)} - \Phi_-^{(\lambda)} \right) \, .
\label{aver}
\end{equation}
Here,  $\Phi^{(\lambda)}$ is the phase field with the harmonics with wavelength shorter than $\Lambda$ removed.
In this standard Kane-Fisher renormalization procedure, the Luttinger parameter is independent of $\lambda$ because a single weak link has no effect on the superfluid far away from its location  (i.e., $dK / d\ln \lambda = 0$, or, in integral form, $K(\lambda)=\mbox{const}$). Equation~(\ref{aver}) means that the effective strength of the weak link flows with
$\lambda$ as
\begin{equation}
J(\lambda) \, =\, J_0 \,  \lambda^{-1/K}  \qquad [J(\lambda) \, \lambda \, \ll \, 1],
\label{J_flow}
\end{equation}
which is  Eq.~(\ref{eq:RGflow_KF}) recast in integral form.

We thus recover the well-known result for fermions/hard-bosons in 1D:  In the thermodynamic limit, a weak link becomes fully transparent for attractive interactions ($K > 1$) and reflects any (even dc) current for repulsive interactions ($K < 1$). However, of importance to us is the flow and the physics at {\it mesoscopic} scales. Specifically, a decisive role in our analysis will be played by a simple fact that, at $K>1$, the  flow (\ref{J_flow}) stops at the wavelength $\lambda_*$ such that $J(\lambda_*) \,\lambda_* ~\sim~1$. We refer to this wavelength as the clutch scale---to underline that the phase field becomes continuous across the link. At length scales $\lambda \gg \lambda_*$, the quantum phase slippages are suppressed, and the link behaves as a classical-field link of strength
\begin{equation}
J_* \, \equiv \, J(\lambda_*) \, \sim \, J_0^{K\over K-1} \; .
\label{J_fstar}
\end{equation}

The Kane-Fisher renormalization of the weak link and its progressively stronger suppression
of the superfluid stiffness is demonstrated numerically in Fig.~\ref{fig:onelink}. It was obtained by a Monte Carlo simulation of the Bose-Hubbard model with parameters corresponding to the superfluid state with $K \approx 3.9$ and where one hopping amplitude was suppressed by a macroscopic factor $2/L$ to form a weak link. It illustrates the most important new ingredient in our theory of the strong-disorder critical point that isolated links are
renormalized by hydrodynamic phonons. This undeniable physics of leak links
is missed in the real-space RG treatment,\cite{Altman} which relies on Coulomb blockade phenomena even in the SF phase.

\section{Prolonged classical flow}
\label{sec:classical}

Before entering the strong-disorder critical regime, superfluid systems with $K(0)\gg 1$ follow a prolonged classical flow \cite{Altman,classical}. In Ref.~\cite{classical} it was argued that the strength of the weakest link in a system of size $L$ scales as a certain power of $L$, conveniently parameterized as $J_0 \sim 1/L^{1-\zeta}$, where $\zeta$ is determined by the microscopic parameters.
However, according to  Eq.~(\ref{J_fstar}), the quantum-renormalized classical theory corresponds to $J_* \sim 1/L^{1-\tilde{\zeta}}$, with
\begin{equation}
\tilde{\zeta} \, = \,   {\zeta K -1 \over K -1}\; .
\label{tilde_zeta}
\end{equation}
Thus, the effective ``classical-field"  exponent $\tilde{\zeta}$ turns out to be a function of $K$.  This fact---central for the scenario revealed below---was overlooked in our Ref.~\onlinecite{classical}.

The microscopic exponent $\zeta$ itself can be measured experimentally/numerically directly by examining the superfluid response of an appropriately large number $N$ of mesoscopic systems of a fixed size $L$. It is expected that the weakest link that can be found under these circumstances has $J_0 \propto 1/N^{1-\zeta}$ because for $\lambda_* > L$ the quantum renormalization
amounts to a constant $N$-independent $L^{-1/K}$ factor, see Eq.~(\ref{J_flow}). For exponentially-rare-exponentially-weak distributions the weakest links are composed of stronger links placed right next to each other
implying that the length of the link $\propto \ln J_0$. This leads to the following requirement on the measurements of $\zeta$: $ \ln N \ll L \ll N$.

On the basis of Eq.~(\ref{tilde_zeta}) we see that the classical-field  approach of Ref.~\onlinecite{classical} only applies in the following two limits: (i) at $\zeta K , \, K \gg 1$, when the quantum renormalization of $\tilde{\zeta}$ is negligible, and (ii) in the superfluid phase beyond the correlation radius corresponding to the saturation of the
superfluid stiffness (and thus $K$) to its infinite-size value.
Most importantly, at the special point $\zeta K =1$  the quantum renormalized $\tilde{\zeta}$ {\it changes sign}. This means that the system can only remain superfluid at $\zeta K \geq 1$.
Moreover, as we show below, the equality indeed corresponds to the critical point,
\begin{equation}
K^{(c)}=1/\zeta \; .
\label{crit_cond}
\end{equation}
As long as $\zeta < 2/3$,  the transition to the Bose glass follows a novel strong-disorder scenario with $K^{(c)}>3/2$.  This {\it inevitably}  happens if $K_0\gg 1$, because the initial (classical) flow requires $\zeta \ll 1$ to suppress superfluidity.

Before proceeding, let us pin-point an explicit contradiction between the real-space RG result \cite{Altman} and the Kane-Fisher renormalization of a single weak link.
The key quantity in real-space RG is the exponent $\alpha$ governing the distribution of renormalized weak links, vanishing
at criticality and taking a small positive value in the SF phase.\cite{Altman, Pielawa} Since real-space RG {\it does not} account for renormalization of isolated links due to long-range zero-point hydrodynamic fluctuations (phonons), we can relate $\alpha$ to $\zeta$ as $\zeta = \alpha / (1 + \alpha)$.
Equation (\ref{crit_cond}) implies then that  a state with small enough but finite $\alpha$ inevitably becomes  incompatible with superfluidity, while the real-space RG puts the critical point at $\alpha=0$.

We can go even a step further. Consider a hypothetical theory, possibly in combination with numerics, capable of producing values $\alpha(L)$ and $K(L)$ up to some finite system size $L$ for the distribution of weak links and the Luttinger parameter, respectively. Renormalization effects by phonons with wavelength larger than $L$ have not been included yet. Consequently, if the criterion $\alpha(L) /[1+\alpha(L)] < 1/K(L)$ is satisfied, the Kane-Fisher renormalization of weak links by long-wave phonons will inevitably result in the insulating state. (Note that $\alpha$ and $K$ can only decrease with the system size.)


\section{Semi-RG flow}
\label{sec:semiRG}
A controlled description of the weak-link quantum criticality is achieved by combining the RG treatment of Ref.~\onlinecite{classical} with the Kane-Fisher renormalization of the link strength at a given length scale. The applicability of hydrodynamics below the correlation radius is guaranteed by the theorem of  critical self-averaging proven in Ref.~\onlinecite{classical} and stating that the superfluid stiffness is well defined for the critical flow as long as it does not vanish in the thermodynamic limit.
If governed by single weak links, the flow of superfluid stiffness $\Lambda_s$ has to obey the equation  $d \Lambda_s^{-1}/d\ln \lambda \propto [J_*(\lambda) \lambda]^{-1}$ (see  Ref.~\onlinecite{classical}). We cast this equation in the form [below $z=\ln(\lambda /\ell_0)$]
\begin{equation}
{d\Lambda_s^{-1}\over dz} \,  \propto\, r(z) \, , \qquad r(z) \, \equiv \,  {\lambda_*\over \lambda} \; ,
\label{flow1}
\end{equation}
by recalling that $J_*(\lambda)$ is the strength of the typical weakest link in a system of size $\lambda$, and $\lambda_* = 1/J_*(\lambda) \equiv \lambda_*(\lambda)$. It is instructive to observe that for $\Lambda_s$ to stay finite in the $z\to \infty$ limit, the ratio $r(z)$ has to obey
the limiting relation
\begin{equation}
\lim_{z\to \infty}z \, r(z) \, =\, 0 \; .
\label{r_limit}
\end{equation}
This relation implies  $\lim_{\lambda \to \infty} \lambda_* / \lambda \to 0$, which is a necessary condition  (but
potentially not a sufficient one because of the occurrence of composite weak links, see below) for single weak
links with the same $\lambda_*$ to be treated as independent. Below we will see that our
semi-RG flow satisfies the condition (\ref{r_limit}).

We now proceed with constructing a self-consistent quantitative description in which
weak links result in a slow flow of $K(\lambda)$ while the flow of $K(\lambda)$ enhances the
renormalization of microscopic weak links. To this end, we recall that the factor $\lambda^{-1/K} $
in (\ref{J_flow}) is, in fact, a product of factors $(\lambda_i/\lambda_{i+1})^{1/K}$
associated with renormalization coming from the wavelength intervals $[\lambda_i,\, \lambda_{i+1}]$.
For a slowly flowing $K(\lambda)$, each term in the product can be written as, $(\lambda_i/\lambda_{i+1})^{1/K(\lambda_i)}$, provided the intervals $[\lambda_i,\, \lambda_{i+1}]$
are small enough to guarantee $K(\lambda_i)\approx K(\lambda_{i+1})$.
This leads to the integral analog of (\ref{J_flow})
\begin{equation}
J(\lambda)  = J_0  \exp\left[ -\int_0^{\ln \lambda} {d\ln \lambda' \over K(\lambda')}\right]  \qquad [J(\lambda) \, \lambda \, \ll \, 1] \, .
\label{J_flow_int}
\end{equation}

Next, we have to generalize Eqs.~(\ref{J_fstar}) and (\ref{tilde_zeta}) for the typical weakest link
at the length scale $\lambda$,
or
$J_0(\lambda)\,  \propto\,  1/\lambda^{1-\zeta}$.
The clutch condition now reads
\begin{equation}
{\lambda_* \over \lambda^{1-\zeta}} \exp\left[ -\int_0^{\ln \lambda_*} {d\ln \lambda' \over K(\lambda')}\right]  \, =\, \mbox{const} \, ,
\label{clutch_1}
\end{equation}
which can  conveniently be  written as
\begin{equation}
z_* +(\zeta -1 )z  -\int_0^{z_*} x(z')\, dz'  \, =\, \mbox{const} \, ,
\label{clutch_2}
\end{equation}
with  $z_*=\ln \lambda_*$ and $x(z)\equiv 1/K(z)$. Differentiating with respect to $z$, we find
\begin{equation}
{dz_*\over dz} \, =\, {1-\zeta \over 1-x(z_*)} \, .
\label{z_*_flow}
\end{equation}

Given that $K(\lambda)=\pi \sqrt{\Lambda_s (\lambda ) \kappa}$ with the $\lambda$-independent compressibility $\kappa$
we see that Eqs.~(\ref{flow1}) and (\ref{z_*_flow}) completely define the semi-RG flow of $K(\lambda)$.
In terms of $x$ and $y=-\ln r \equiv z - z_*$, we get
 [below $x_0\equiv x(z=0)$]
\begin{equation}
 \frac{1}{x^2_0} {dx^2\over dz}  \, =\, e^{-y} \; ,
\label{RG1}
\end{equation}
\begin{equation}
{dy\over dz } \, =\,  \frac{\zeta-x}{1-x} \;.
\label{RG2}
\end{equation}
By Eq.~(\ref{z_*_flow}),  $x$ in the r.h.s. of (\ref{RG2}) should be understood as $x\equiv x(z-y)$.
However, we are allowed to substitute $x(z-y) \to x(z)$
because by Eq.~(\ref{RG1}) the correction is supposed to be small, $(ydx/dz)/x \ll 1$.
[And on the critical line, $y/z \to 0$ at $z\to \infty$, as we will see soon.]
As can readily be verified, the classical flow equations from Ref.~\onlinecite{classical} are recovered by ignoring the $x$ dependence in Eq.~(\ref{RG2}). The Kane-Fisher equation is recovered by ignoring the $(\zeta - x)$ dependence in Eq.~(\ref{RG2}), which is easiest seen by ignoring the second term in Eq.~(\ref{clutch_2}).

Note that $\zeta$ is a bare parameter not subject to the RG flow (hence the name semi-RG flow) because we are dealing with single links and ignore pairs of links (see below). Moreover, the Luttinger parameter is the key quantity governing the flow toward an insulating state. Both facts are in sharp contrast with the real-space RG treatment,\cite{Altman} where Coulomb blockade physics plays the key role instead of hydrodynamics and where single weak links cannot be treated along the lines of Kane-Fisher because the real-space RG phenomenology is such that single weak links surrounded by a Luttinger cannot occur in a context of a strong randomness.

The first integral of equations (\ref{RG1})--(\ref{RG2}) can easily be found in a closed form,
\begin{equation}
 (1-\zeta)[x+\ln(1-x)]+x^2/2 \, =\, -(x^2_0/2)e^{-y}+C   \;.
\label{eq4b}
\end{equation}
Since $y(\lambda \to \infty) \to \infty$, the value of $C$ can be expressed as
\begin{equation}
C=(1-\zeta)[x_{\infty} + \ln(1-x_{\infty})] + x_{\infty}^2/2 \;.
\label{eq4c}
\end{equation}
The critical point for the SF-BG transition is located at $x_\infty=1/K(\infty) = \zeta$ and corresponds
to the strong-disorder scenario if $x_\infty < 2/3$. At criticality,
$C=\zeta-\zeta^2/2+(1-\zeta)\ln(1-\zeta)$.
One can further see that
\begin{equation}
x(z)\, =\,   \zeta -2(1-\zeta)/z\qquad  \mbox{(at criticality)} \, ,
\label{criticality}
\end{equation}
implying, in particular, the critical behavior
$r(z)\propto 1/z^2$, consistent with Eq.~(\ref{r_limit}).

Finally, the dependence of $x_{\infty}$ on the external parameters
is of the KT-type, $x_{\infty}(g)=\zeta-D\sqrt{g-g_c}$, implying the standard KT-type exponential divergence
of the correlation length on approach to the critical point. The easiest way to derive these results
is to approach the critical point along the $C=const$ trajectory and consider $\zeta -\zeta_C$ and $\zeta-x$
as small parameters to simplify the equations.

\begin{figure}[htbp]
\includegraphics[scale=0.4,angle=0,width=0.8\columnwidth]{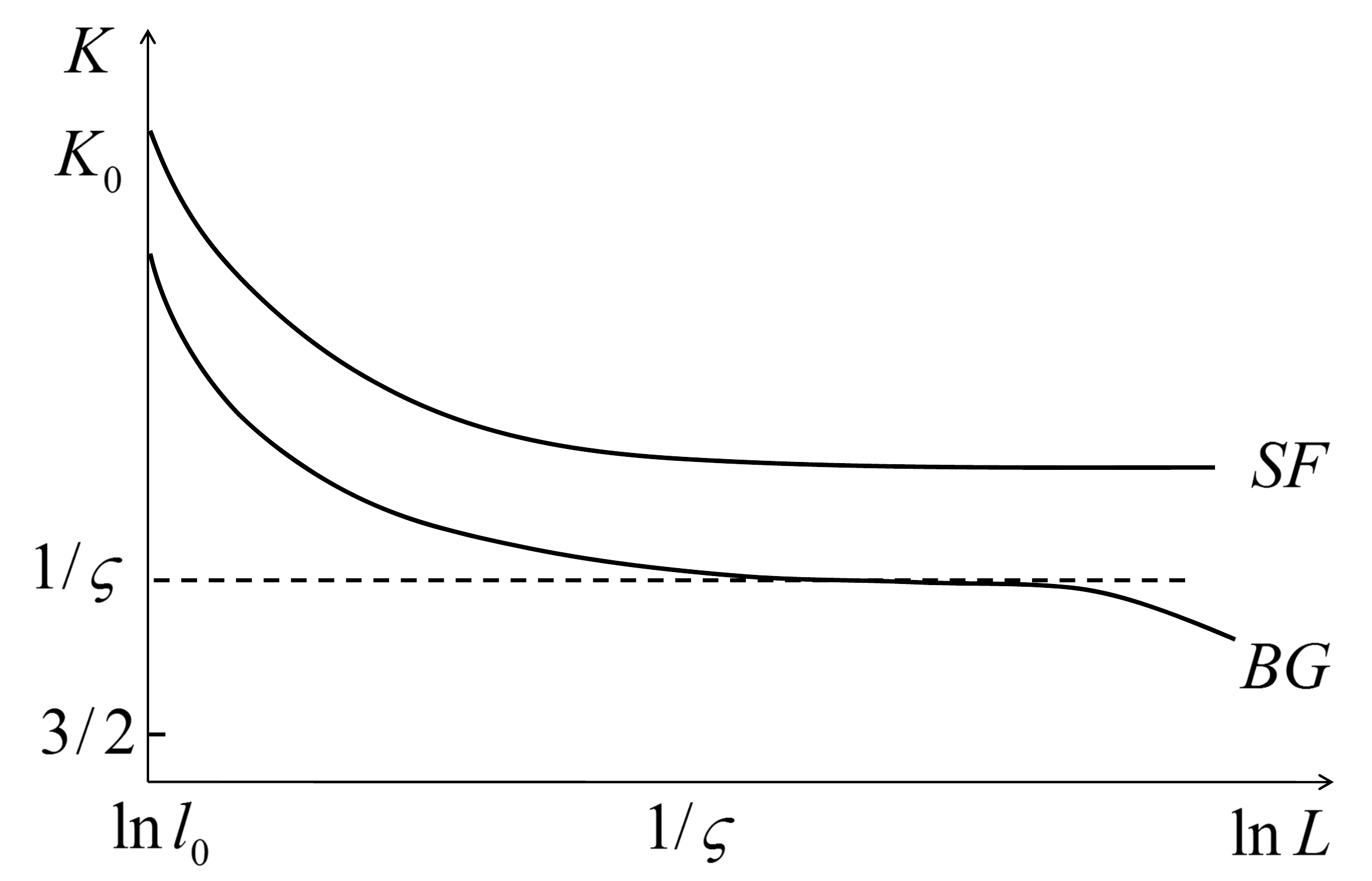}
\caption{\label{fig:1} Flow of the Luttinger parameter with system size in the vicinity of the critical point
when we have $K_0 \gg 1$ at the UV cutoff scale $\ell_0$. The initial behavior of the flow pertains up to the length scale $1/\zeta$ [see Eq.~(\ref{eq5c})], when $K$ will level off to a constant in the superfluid phase (SF). If the flow reaches the condition
$K(L)=1/\zeta \sim K_0^{2/3}$, then $K$ will further decrease to 0 and the thermodynamic phase is a Bose glass (BG). }
\end{figure}
In the limit $K(\ell_0)\gg 1$ (i.e., $x_0 \ll 1$) and $\zeta  \ll 1$, but $\zeta > x_0$,
the solution is particularly simple:
\begin{equation}
\zeta \frac{x^2}{2}-\frac{x^3}{3} \, =\,  -\frac{x_0^2}{2}\, e^{-y}+C\;, \quad
C=\zeta \frac{x_{\infty}^2}{2}-\frac{x_{\infty}^3}{3}\;.
\label{eq5c}
\end{equation}
At the SF-BG criticality, when $x_{\infty}=\zeta$, we have $C=\zeta^3/6$.
We further notice that the case $(x_0,x) \ll \zeta \ll 1$ reduces to the
classical RG flow because $x$ can be neglected in the r.h.s.
of $dy/dz$. Then
\begin{equation}
x(L)\, =\, x_0 \sqrt{1 + e^{-y_0}(1-e^{-\zeta z} )/\zeta } \;,
\label{eq5c}
\end{equation}
with
\begin{equation}
x_{\infty}\, =\, x_0 \sqrt{1 + e^{-y_0}/\zeta }\, \propto\,  \frac{x_0}{\sqrt{\zeta}} \;.
\label{eq5d}
\end{equation}
From this estimate we see that in the classical field limit the SF-BG transition
corresponding to $x_{\infty}=\zeta$ occurs when $x_{\infty} \propto x_0^{2/3}$.
The overall picture is illustrated in Fig.~\ref{fig:1}.

\section{Irrelevance of Coulomb blockade physics}
\label{sec:coulomb}


Consider pairs of weak links separated by a large distance $d$---called $d$-pairs for brevity,
with the number of particles on the $d$-interval being well defined (the Coulomb blockade regime).
Let us show that renormalization of $\Lambda_s$ by such complexes can be neglected.
To this end, we first establish the functional form of the Kane-Fisher factor [see Eq.~(\ref{J_flow_int})]
in the asymptotic $\lambda \to \infty$ limit by using Eq.~(\ref{criticality})
\begin{equation}
f(\lambda ) = \exp \left[-\int_0^{\ln \lambda} x(z)\, dz \right] \propto \lambda^{-\zeta} \ln^{2(1-\zeta)}(\lambda) \; .
\label{RGfactor}
\end{equation}

Consider now some length scale $\lambda$ and account for the $d$-pairs
which have a probability of the order of unity to occur at this scale
(pairs with higher density are absorbed into the renormalized value of
$\Lambda_s (\lambda)$; unlikely events will be accounted for at larger scales).
For clarity, we start with the case of two weak links having similar values of
$J_0$ separated by a distance scale $d$. The requirement for such pairs to occur
with a probability of order unity,
\begin{equation}
 \left[ J_0^{1/(1-\zeta)}\right]^2 d = 1/\lambda  \qquad \mbox{(for $d$-pairs)} \; ,
\label{probability1}
\end{equation}
translates into the $J_0^2 =(1/d\lambda)^{1-\zeta}$ relation for the strength
of weak links in the pair. As long as $J(d)=J_0f(d)$ remains smaller than the
``charging" energy of the system's interval between the links, $\kappa /d$, one can
use  the result of second-order perturbation theory to estimate the strength
of the composite link as $J(d)^2/(\kappa /d)$. The contribution of the pair to the
renormalization of $\Lambda_s^{-1}$ is given, as before, by the ratio
$\lambda_*/\lambda$ where $\lambda_*$ is defined by
\begin{equation}
\left[J_0^2 f^2(d) d \right]\: \frac{f(\lambda_*)}{f(d)} \: \lambda_* \sim 1 \qquad \mbox{(for $d$-pairs)} \; .
\label{pairclutch}
\end{equation}
The second factor accounts for the composite link renormalization between the $d-$ and
$\lambda_*$-scales. By substituting here Eqs.~(\ref{probability1}) and (\ref{RGfactor})
we readily find
\begin{equation}
 \left[ \ln^2(d) \ln^2(\lambda_*) \; \frac{\lambda_*}{\lambda} \right]^{1-\zeta} \sim 1 \qquad \mbox{(for $d$-pairs)} \; ,
\label{pairdone}
\end{equation}
or, by replacing $\lambda_*$ with $\lambda$ up to  logarithmic precision,
\begin{equation}
\frac{\lambda_*}{\lambda} \sim \frac{1}{ \ln^2(d) \ln^2(\lambda)} \qquad \mbox{(for $d$-pairs)} \; .
\label{pairdone2}
\end{equation}

We immediately see that the contribution of large $d$-pairs is suppressed
by a factor $\ln^{-2}(d)$. Most importantly, the integral over the pair scales
$\int d\ln (d) $ is converging at the lower limit where microscopic pairs
(and other multi-link complexes) are part of the original
exponentially-rare-exponentially-weak distribution of single links.
%
%
The same final conclusion (\ref{pairdone2})
is reached for a pair of links with different strength, $J_1=J_0\delta$, $J_2=J_0/\delta $.
Since $J_1J_2=J_0^2$, all equations in the analysis presented above
remain identically the same.

\section{Conclusion and Outlook}
\label{sec:conclusion}

We presented an asymptotically exact theory for the SF-BG transition in the presence of appropriately strong disorder by combining the classical field flow of Ref.~\onlinecite{classical} with the Kane-Fisher renormalization, originally derived for single weak links.\cite{KaneFisher} This constitutes the crucial difference with the real-space RG treatment introduced by Altman and coworkers, as the latter is unable to address the Kane-Fisher renormalization---since it does not treat properly the phonon degrees of freedom in the system.
The hallmark of our theory is the relation $K^{(c)} = 1/\zeta$ stating that there are no superfluids with a Luttinger parameter smaller than $1/\zeta$; all future work has to deal with this microscopic quantity.

We have checked that the available data from Refs.~\onlinecite{classical, Pielawa} are compatible with the present scenario (and, in particular, with a Kosterlitz-Thouless-type transition at a non-universal value of $K_c$), but the data are insufficient for studying the transition accurately and/or extracting $\zeta$.

Our treatment applies to both Bose and Mott glasses. In the latter case, the system remains
compressible at criticality even though it is incompressible on the insulating side (the renormalization
of $\kappa$ starts when $K$ drops to values close to $2$).
The semi-RG equations that we derived here can straightforwardly be upgraded
to a system of three equations describing both the strong-disorder
and Giamarchi-Schultz (or Mott-glass) criticalities, as well as the competition between the two.
This is achieved by accounting for the standard (for KT theory)
instanton--anti-instanton renormalization terms in the flows of
$\Lambda_s$ (and $\kappa$, if necessary), and introducing the RG equation for the flow
of concentration of the instanton--anti-instanton pairs (for the details, see Ref.~\onlinecite{Kashurnikov96}).

Furthermore, in the Mott glass case (where instantons have no phase factors), the ground-state 1D quantum system is directly mapped onto a
2D  finite-temperature ``scratched-disordered" superfluid film. The film is supposed to have a peculiar disorder in the form of straight parallel scratches cutting  through the film.
If the disorder is strong enough to guarantee $\zeta < 1/2$, the film experiences the superfluid-to-normal phase transition of the above-discussed strong-disorder universality class, happening at the critical
temperature $T_c=\pi \zeta \hbar^2 n_s/m$ (with $n_s$ the superfluid density and $m$ the mass of the atoms), thus preempting the usual  Berezinskii-Kosterlitz-Thouless transition.
Despite a different transition scenario, the state at $T>T_c$ is just a normal film. 
This is seen from the fact that the vortex pairs are {\it dangerously} irrelevant with respect to
the scratches: When $n_s$ is suppressed to zero at $T>T_c$, the vortex-antivortex pairs inevitably 
proliferate, rendering the final phase indistinguishable from the standard normal state.
Likewise, there are no two different Bose glass phases because the rare weak links and the instanton--anti-instanton pairs are dangerously irrelevant with respect to each other. 
Hence, while only one of the two is responsible for criticality, the other one also becomes important on the insulating side (at distances at which the value of $K$ becomes appropriately small), removing the potential qualitative difference between the glasses.

We are grateful to Ehud Altman, Thierry Giamarchi, Susanne Pielawa, and Anatoli Polkovnikov for valuable discussions and S. Pielawa for sharing her numerical data for the Luttinger parameter with us. This work was supported by the National Science Foundation under the grant PHY-1314735, FP7/Marie-Curie Grant No. 321918 (``FDIAGMC"),  FP7/ERC Starting Grant No. 306897 (``QUSIMGAS") and by a grant from the Army Research Office with funding from DARPA.

\end{document}